# A Novel Metric for Assessing Climatological Surface Habitability

Hannah L. Woodward[1,2], Andrew J. Rushby[1,2], and Nathan J. Mayne[3]
[1] School of Natural Sciences, Birkbeck, University of London, Malet Street, London, WC1E 7HX, UK; hannah@woodward.sh, a.rushby@bbk.ac.uk
[2] Centre for Planetary Sciences, University College London, Gower Street, London, WC1E 6BT, UK
[3] Department of Physics and Astronomy, University of Exeter, Stocker Road, Exeter, Devon, EX4 4SB, UK; n.j.mayne@exeter.ac.uk

## Abstract

Planetary surface habitability has so far been considered, in the main, on a global scale. The increasing number of 3D modelling studies of (exo)planetary climate has highlighted the need for a more nuanced understanding of surface habitability. Using satellite-derived data of photosynthetic life to represent the observed surface habitability of modern Earth, we validate a set of climatologically defined metrics previously used in habitability studies. The comparison finds that the metrics defined by surface temperature alone show spatial patterns of habitability distinct to those defined by aridity or water availability, with no metric able to completely replicate the observed habitability. We build upon these results to introduce a new metric defined by the observed thermal limits of modern Earth-based life, along with surface water fluxes as an analog for water and nutrient availability. Furthermore, we pay attention to not only the thermal bounds of macroscopic complex life, but additionally the limits of microbial life which have been vital to the generation of Earth's biosignatures, thus expanding considerations of climatic habitability out of a historically binary definition. Repeating the validation for our metric, along with another that uses a similar definition that incorporates conditions for both temperature and water availability, shows a significant improvement in capturing the broad patterns of surface habitability, laying the groundwork for more comprehensive assessments of potential life-supporting climates upon other planets.

*Unified Astronomy Thesaurus concepts:* Astrobiology (74); Planetary climates (2184); Habitable planets (695); Earth (planet) (439); Planetary atmospheres (1244)

## 1. Introduction

Habitability has been investigated across many spatial and temporal scales, from the Galactic habitable zone (G. Gonzalez 2001; C. H. Lineweaver et al. 2004) to the ecological niches of extremophiles (N. Merino et al. 2019; L. Ratliff et al. 2023), as well as the dependence upon stellar age and (bio)geochemical cycling (A. J. Rushby et al. 2013, 2018). The definition of the habitable zone, that is the circumstellar region in which the global mean surface temperature of a planet permits the persistence of liquid water, initially enabled a binary view of whether a planet, as a whole, may be habitable (J. F. Kasting et al. 1993; R. K. Kopparapu et al. 2013). The increasing number in the use of climate models, including energy balance models (EBMs) and 3D general circulation models (GCMs), to study (exo)planetary climate has further introduced a variety of definitions for sub-planetary-scale surface habitability: temperature ranges optimal for complex or human life (e.g., L. Silva et al. 2016, 2017; A. H. Lobo et al. 2023), ice-free fraction (Y. Hu & J. Yang 2013; M. Turbet et al. 2016; A. D. D. Genio et al. 2019; J. Yang et al. 2020; D. E. Sergeev et al. 2022), temperature ranges in which liquid water may persist (D. S. Spiegel et al. 2008; G. Vladilo et al. 2013; T. Jansen et al. 2019), as well as of water availability through the consideration of water fluxes on land (A. D. Del Genio et al. 2019). Some more recent definitions have become more sophisticated by using a combination of climatological variables to define and investigate surface habitability across a range of orbital and planetary configurations across Earth and Earth-like planets (F. He et al. 2022; A. Farnsworth et al. 2023; A. D. Adams et al. 2025). However, none of these definitions have yet been validated against how life is observed to be spatially distributed across Earth. With recent detections of potentially habitable exoplanets (e.g., R. Luger et al. 2017; D. Kossakowski et al. 2023) and advances in the characterization of rocky planet atmospheres (e.g., A. B. Claringbold et al. 2023; A. P. Lincowski et al. 2023; J. Lustig-Yaeger et al. 2023; S. E. Moran et al. 2023; S. Zieba et al. 2023), it should therefore be a priority to consider how the use of GCMs can best assist the observational community in the search for potentially habitable planets. In this way, we aim to not only review a variety of climatologically based metrics of surface habitability with respect to modern Earth-based surface life, but additionally define a new metric of surface habitability which more closely reflects the spatial patterns of life observed on Earth.

On Earth, life exists and thrives across a multitude of habitats with varying physicochemical conditions, including temperature, pH, pressure, salinity, and radiation exposure (N. Merino et al. 2019; E. M. Sogin et al. 2021). On the surface, complex multicellular life tends to dwell and thrive in the more temperate regions of Earth, while some life, particularly microbial, is able to thrive over a wider range of temperatures spanning from –20 to 122°C (A. Anav et al. 2015; L. Silva et al. 2016; N. Merino et al. 2019). Common to all known lifeforms is a requirement for liquid water, as a solvent, along with sources of energy and nutrients (C. Cockell et al. 2016). While necessary, the presence of liquid water alone is not sufficient to imply an abundance of life; such an example is seen in the oligotrophic gyres of the Earth's oceans, which are regions of downwelling that are nutrient limited and thus only able to support low biological productivity (C. R. McClain et al. 2004; E. G. Jones & C. H. Lineweaver 2010). Similarly, it is possible for life to exist in frozen conditions; taking, for example, algal blooms observed within and underneath sea ice, the evolution of antifreeze proteins within fish, or microbes within cryoconite or brine pockets (G. L. Fletcher et al. 2001; A. Clarke 2003; P. B. Price 2007; J. Cook et al. 2016; M. Ardyna et al. 2020). These examples not only highlight the extraordinary adaptive capabilities of life but also the importance of nutrients in considerations of habitability (C. Cockell et al. 2016; A. E. Nicholson et al. 2022). The anisotropic distribution of nutrients across the Earth is, in part, determined by atmospheric-driven processes (e.g., aeolian transport and oceanic upwelling), along with surface processes such as chemical weathering and outgassing (J. C. G. Walker et al. 1981; R. T. Pierrehumbert 2002; A. J. West et al. 2005; A. J. Rushby et al. 2018). In this way, the spatial patterns of surface habitability may also be tied to the atmospheric dynamics, and in particular, the structure of the atmospheric cells, which in turn are determined by planetary-scale properties including rotation rate, stellar flux, and atmospheric composition (A. D. Del Genio & R. J. Suozzo 1987; A. P. Showman et al. 2013; Y. Kaspi & A. P. Showman 2015; J. Haqq-Misra et al. 2018; T. Jansen et al. 2019; M. Hammond & N. T. Lewis 2021; D. A. Williams et al. 2024; A. D. Adams et al. 2025).

Many studies of extraterrestrial surface habitability apply a binary assignment of habitable or uninhabitable, yet their definitions are based upon assumptions that do not completely encompass the factors discussed above (e.g., D. S. Spiegel et al. 2008; T. Jansen et al. 2019; D. E. Sergeev et al. 2022; A. H. Lobo et al. 2023). Although life is indeed less prevalent in more extreme environments, it is important to consider that the greatest contributor to Earth's main current biosignature, atmospheric oxygen, is plausibly attributable to photosynthetic bacteria during the Great Oxygenation Event (P. W. Crockford et al. 2023). Furthermore, microbial life has not only dominated for the majority of the biological history of Earth, but is also responsible for approximately half of global primary production, that is, the fixation of carbon into organic material through photosynthesis (C. B. Field et al. 1998; P. W. Crockford et al. 2023). On the other hand, the Earth's climate is being significantly impacted by human life despite humans accounting for a mere ≈1% of global biomass, including observed changes in the atmospheric composition as well as the production of potentially detectable technosignatures (Y. M. Bar-On et al. 2018; IPCC 2021; J. Haqq-Misra et al. 2022; S. Seager et al. 2023). Investigations of surface habitability should therefore include the range of climatological conditions within which both microbial and complex life can flourish, and in particular, to the extent such that its existence may be detectable (A. E. Nicholson & N. J. Mayne 2023).

Putting these considerations together in tandem with a comparison against other similarly defined metrics, we propose a novel climatologically based metric of surface habitability, which explicitly considers the temperature ranges required by current known life, as well as establishing a condition of "water availability" through the incorporation of surface water fluxes. Furthermore, the impact of nutrient availability is implicitly included within the metric definition through the acknowledgment that both water and nutrients are influenced by similar atmospheric-driven processes such as weathering and wind-driven transport. Applied to 2D geospatial data of a planetary surface, the metric allows an examination into the broad spatial patterns of surface habitability as well as being reduced into a global habitable fraction. This offers the opportunity to examine snapshots of a mean planetary climate and investigate how both spatial patterns and fractional habitability may evolve over geological timescales on a single planet, as well as providing a framework for intercomparisons of surface habitability across model or planetary configurations (e.g., J. T. O'Malley-James et al. 2014; J. Yang et al. 2019; T. J. Fauchez et al. 2020; M. J. Way et al. 2021; A. Farnsworth et al. 2023; A. D. Adams et al. 2025). Furthermore, the metric distinguishes between areas which harbor climatological conditions more suitable for microbial and macroscopic complex life as we see on modern Earth. Future applications of the metric may be, for example, to investigate the implications upon the generation of detectable biosignatures and/or technosignatures, for example through the coupling with or forcing of models which explore the biosphere–climate interaction (S. Seager 2013; A. E. Nicholson & N. J. Mayne 2023; R. Arthur et al. 2024). Along with using Earth as a testbed for the metric, this also leads to the potential of exploring *superhabitable* conditions, in which a planet may be considered to be more habitable than Earth, for example through a higher habitable fraction (D. Schulze-Makuch et al. 2020).

The definition of our surface habitability metric is presented in Section 2, followed by an outline of the data and methods in Section 3. A selection of surface habitability metrics, including our newly defined metric, are then validated against the distribution of photosynthetic life on Earth in Section 4. Concluding remarks are given in Section 5.

## 2. Metric Description

Our habitability metric ($H$) is defined using three climatological variables: surface air temperature $T_s$ (in kelvin), precipitation $P$ (in mm $yr_{\oplus}^{-1}$), and evaporation $E$ (in mm $yr_{\oplus}^{-1}$). As the metric aims to consider climates optimal for both autotrophic and heterotrophic life, surface temperature was selected over variables such as UV flux or photosynthetically active radiation, which would discriminate toward specific pigments (such as chlorophyll-a; Chl-*a* hereafter) or biochemical reactions, which may be optimized to the solar spectral energy distribution, as well as more generally toward photosynthetic life over chemosynthetic life, which, for example, may thrive on the nightside of tidally locked planets (N. Y. Kiang et al. 2007; O. R. Lehmer et al. 2021; E. M. Sogin et al. 2021; S. Chitnavis et al. 2024). First, two temperature ranges are defined that represent the thermal limits for *microbial* and *complex* life, respectively. The microbial range is based upon the observed limits of extremophilic microbial life to be metabolically active and able to reproduce (E. M. Rivkina et al. 2000; K. Takai et al. 2008; N. Merino et al. 2019). We adopt the complex range from the "complex life" habitability metric defined by L. Silva et al. (2016), which is a conservative estimate based upon the temperature ranges inhabited by most multicellular poikilotherms, or in other words, organisms whose internal body temperature directly depends on and varies with the ambient environmental temperature. Modern poikilotherms include most plants, invertebrates, and ectothermic vertebrates (such as reptiles, amphibians, and fish), and represent both the most common and evolutionarily ancient form of complex life on Earth. Homeotherms, which include mammals and birds, were not explicitly included within the metric definition as their ability to maintain a constant internal temperature distinct to the ambient temperature makes it more difficult to constrain their environmental limits (A. Clarke 2014; L. Silva et al. 2016). However, taking into consideration that poikilotherms form an evolutionary precursor to homeotherms as well as some of the shared biological mechanisms that set the outlined thermal limits, L. Silva et al. (2016) argue that their metric remains generalizable to most forms of macroscopic complex life, including potential "Earth-like" extraterrestrial life. Putting the two temperature ranges together, we define the habitable temperature condition $H_T(T_s)$ as

$$H_T(T_s) = \begin{cases} \text{complex} & \text{if} \quad 273.15 \leqslant T_s \leqslant 323.15 \text{ K}, \\ \text{microbial} & \text{if} \quad 253.15 \leqslant T_s \leqslant 395.15 \text{ K}, \\ \text{limited} & \text{otherwise}. \end{cases} \tag{1}$$

Although the microbial temperature range in Equation (1) includes water outside of the liquid state at 1 bar pressure, retaining this range acknowledges both the fully observed limits of life while also allowing investigations into other environments, for example different atmospheric pressures or salinity, in which water may still exist in liquid form outside of 0–100°C. The second condition for habitability is based upon the requirement for liquid water (C. Cockell et al. 2016), which we extend to a general condition of "water availability" through the consideration of surface water fluxes as represented by precipitation $P$ minus evaporation $E$, along with a minimum precipitation rate equivalent to the definition of a desert on Earth (I. Noy-Meir 1973):

$$H(T_s, P, E) = \begin{cases} H_T(T_s) & \text{if} \quad P - E \geqslant 0 \text{ and } P \geqslant 250 \text{ mm yr}_{\oplus}^{-1}, \\ \text{limited} & \text{otherwise}. \end{cases} \tag{2}$$

While the concept of water availability is not new to terrestrial investigations of primary productivity (R. R. Nemani et al. 2003), it may at first seem an unusual condition to apply to the oceans, which are obviously abundant in water. However, we highlight here that the broad spatial patterns of $P$ and $E$ are primarily controlled by the global atmospheric circulation and as such are related to other variables such as surface wind, which in turn influence the transport (whether directly via the wind itself or indirectly through wind-driven oceanic upwelling, for example) and subsequent availability of nutrients (J. C. G. Walker et al. 1981; R. T. Pierrehumbert 2002; A. P. Showman et al. 2013). Put together, Equation (2) is a binary requirement for a habitable environment, while Equation (1) controls the assignment of microbial versus complex habitability. If either Equation (1) or Equation (2) is not satisfied, then the habitability category is assigned as *limited*. We emphasize that these categories are nonstrict in the sense that they provide an indication of the predominant form (or absence) of life, so for example, it is possible that some complex life may inhabit areas designated as microbial or limited habitability. Additionally, it is important to note that complex habitability is a subset of microbial habitability, thus, an environment marked as complex habitability should be inhabited by both complex and microbial life. For studies of hypothetical habitability on "water-limited" worlds (e.g., M. Turbet et al. 2022; A. H. Lobo et al. 2023), it may be more appropriate to only apply the temperature condition (Equation (1)).

## 3. Data and Methods

### 3.1. Data

#### 3.1.1. Earth Climatological Data

All metrics involved in the comparison make use of the ERA5 global monthly reanalysis (H. Hersbach et al. 2023) to calculate the 2D surface habitability of Earth. ERA5 lies on a $0.^{\circ}25 \times 0.^{\circ}25$ latitude–longitude grid and although it shows some bias in precipitation over the tropics is considered to perform well in representing global surface climate in comparison to observation and other reanalyses (H. Hersbach et al. 2020, 2023; B. Hassler & A. Lauer 2021). In particular, the variables used among the habitability metrics include surface air temperature ($T_s$), evaporation ($E$), potential evapotranspiration (PET), precipitation ($P$), and sea ice concentration (SIC). PET in ERA5 is calculated using the assumptions of "crops/mixed farming" vegetation and "no stress from soil moisture".[4]

#### 3.1.2. Earth-observed Habitability Data

While the metrics make use of climatological variables to predict surface habitability, we look for a data set that directly measures the presence of life to represent the observed habitability, taking the assumption that an abundance of life constitutes a habitable environment and conversely a scarcity of life indicates an environment of limited habitability. In particular, photosynthetic life was chosen as an analog for the abundance and distribution of surface life on Earth for three main reasons: first, photosynthetic life is the dominant form of autotrophic life. Plants in particular contribute up to 80% of global biomass, although it is important to highlight that biomass is not necessarily a signifier of photosynthetic activity; marine primary producers only account for 20% of total marine biomass but remain the main contributor in sustaining higher trophic levels, as well as contributing almost equally toward global net primary production as terrestrial primary producers (C. B. Field et al. 1998; Y. M. Bar-On & R. Milo 2019). Second, the pigment Chl-*a*, used by most photosynthetic life, produces a unique reflectance spectrum in which remote-sensing techniques are able to generate global-scale data sets that are representative of photosynthetic productivity (e.g., S. Sathyendranath et al. 2019; K. Didan 2021). Finally, as primary producers, we assume that where photosynthetic life is able to thrive, so too can heterotrophic life.

Since there is no single global data set of photosynthetic life, the monthly Moderate Resolution Imaging Spectroradiometer (MODIS) Terra data product of normalized difference vegetation index (NDVI; K. Didan 2021) and the Ocean Color Climate Change Initiative (OC-CCI) v6.0 data product of monthly mean surface Chl-*a* concentration (S. Sathyendranath et al. 2019) were selected to represent terrestrial and marine life, respectively. MODIS NDVI lies on a 9 km latitude–longitude grid and represents vegetation density and health through the measured difference in spectral reflectance between the visible red (Red) and near-infrared (NIR) as (NIR − Red)/(NIR + Red). The calculation of NDVI exploits the terrestrial vegetation "red edge," that is, the strong contrast between these two spectral bands which arises due to the photosynthetic pigment (such as chlorophyll) showing strong absorption across the visible red wavelengths while the leaf structure exhibits strong reflectance in the NIR (S. Seager et al. 2005; N. Pettorelli 2013). As a ratio of spectral reflectance values, NDVI is unitless and ranges from –1 to +1. OC-CCI Chl-*a* concentration is an ocean color product representing the abundance of phytoplankton, in mg m$^{-3}$ and is calculated through the measured ratio of reflected blue to green light. It lies on a 4 km latitude–longitude grid and is an aggregate of data from multiple satellites that is validated against in situ data (S. Sathyendranath et al. 2019). Due to the high viewing angle and presence of sea ice, the product lacks data for latitudes exceeding 50° during the respective hemispheric winter months. To account for this issue, the Chl-*a* concentration monthly climatology was linearly interpolated across the months with missing data, and the resulting interpolated values weighted by the respective SIC sourced from the Hadley Centre Sea Ice and Sea Surface Temperature (HadISST) reanalysis (N. A. Rayner et al. 2003; H. A. Titchner & N. A. Rayner 2014). A minimum value of Chl-*a* concentration was set to 0.01 mg m$^{-3}$, in line with observed underice values (A. Randelhoff et al. 2020; M. Hague & M. Vichi 2021). The HadISST data set incorporates both satellite and in situ data which is then interpolated to provide spatially continuous monthly data at a 1° × 1° resolution on a latitude–longitude grid.

[4] Sourced from the European Centre for Medium-Range Weather Forecasts (ECMWF) Parameter Database (https://codes.ecmwf.int/grib/param-db/228251), noting that the PET variable in ERA5 is named potential evaporation.

These remotely sensed variables were chosen so as to avoid any further introduction of bias from modeled variables such as primary productivity or biomass, while retaining a strong correlation to the density, distribution, and productivity of photosynthetic life (K. S. Johnson & M. B. Bif 2021; T. J. Ryan-Keogh et al. 2023). With this in mind, both products are susceptible to some bias, mostly at higher values where the signal becomes saturated (e.g., A. Huete et al. 2002; V. S. Saba et al. 2011), although this is not problematic for our validation, which sets thresholds for microbial and complex habitability at low to intermediate values of NDVI and Chl-*a* concentration. In addition to the seasonal high-latitude data gaps discussed above, OC-CCI Chl-*a* concentration additionally shows a positive bias for values less than 0.3 mg m$^{-3}$, which increases in magnitude with decreasing Chl-*a* concentration.[5]

### 3.2. Methods

#### 3.2.1. Data Preprocessing and Metric Habitability

Each data set (ERA5, NDVI, and OC-CCI Chl-*a* concentration) was regridded to a 1° × 1° latitude–longitude grid, to match the data set with the coarsest resolution, HadISST SIC. Land and sea masks were applied to the terrestrial (NDVI) and marine (OC-CCI Chl-*a* concentration) observational data, respectively. Monthly mean climatologies across 2003–2018 were first calculated, with the year range selected for maximum temporal overlap across data sets, as well as over a long-enough period to reduce the influence on climate from interannual climate variability. At this stage, the gap filling for OC-CCI Chl-*a* concentration was performed, as well as the calculation of aridity index (AI) used in the A. D. Del Genio et al. (2019) metrics. The AI represents the dryness of an environment through a balance of surface water fluxes and is defined as

$$\mathrm{AI} = \frac{P}{P + \mathrm{PET}}, \quad (3)$$

for precipitation $P$ and potential evapotranspiration PET, with values ranging from 0 to 1 and lower values representing a more arid environment (A. D. Del Genio et al. 2019). PET represents the maximum rate of water loss from the surface to the atmosphere through the processes of evaporation and transpiration assuming an unlimited water supply. Its calculation typically incorporates climatological variables such as surface radiative fluxes, air temperature, saturation vapor pressure, humidity, and wind speed, along with variables relating to the vegetation such as land cover type and fraction, leaf area index, canopy height, or stomatal resistance (H. L. Penman & B. A. Keen 1948; C. H. B. Priestley & R. J. Taylor 1972; J. L. Monteith 1981; J. Scheff & D. M. W. Frierson 2015). For a more comprehensive overview of both PET and AI, we refer the reader to A. D. Del Genio et al. (2019). Finally, an annual mean climate across this period was calculated for all variables from the monthly climatologies.

A set of climatologically defined metrics that have been used previously in studies of broad-scale surface habitability were selected for the comparison, with resolutions ranging from 1° latitudinal belts in a 1D EBM to a 4° × 5° latitude–longitude grid in a GCM (Table 1). In this way, all metrics in the comparison aim to assess habitability at similarly coarse resolutions and not at particular fine-scale environmental niches such as a hot spring or soda lake (e.g., N. Merino et al. 2019). The selection includes definitions of habitable surface temperatures ($H^{\mathrm{S08}}$, $H^{\mathrm{S16}}$), ice-free area ($H^{\mathrm{IF}}$), water availability ($H^{\mathrm{DG19H}}$, $H^{\mathrm{DG19NA}}$), as well as a combination of climate variables ($H^{\mathrm{W25}}$, $H^{\mathrm{A25}}$). We excluded more complicated definitions such as that used to investigate mammal habitability of future Earth (A. Farnsworth et al. 2023), which was explicitly reliant upon seasonality along with a vegetation component within the GCM to calculate bare soil fraction, and as such, not a definition easily generalized to other planets. The 2D mean climate surface habitability was calculated for each metric outlined in Table 1. Although in its original context $H^{\mathrm{A25}}$ was only applied to the terrestrial domain (F. He et al. 2022; A. D. Adams et al. 2025), its definition, which uses $T_s$ and $P$, is also applicable to the marine domain. We have therefore extended its application to the full global domain for a more comprehensive comparison. For each of the microbial and complex categories, the resultant fractional habitability $f_H$ was calculated as the weighted fraction of all grid cells which satisfy the respective habitability conditions, as introduced by D. S. Spiegel et al. (2008) and expanded into two dimensions by T. Jansen et al. (2019). To aid comparison, where previous metric definitions have relied upon a binary assignment of habitability, unless otherwise specified by the respective definition, a *habitable* assignment is here relabeled as microbial. In addition to examining global habitability where possible, we also inspect the habitability of marine and terrestrial domains individually, enabling a comparison of metrics only applicable to either of these domains.

[5] Source: OC-CCI Ocean Color Product User Guide for v6.0 (https://docs.pml.space/share/s/fzNSPb4aQaSDvO7xBNOCIw).

**Table 1**
Overview of Metrics Used in the Comparison of Surface Habitability on Earth with the Conditions Required for Complex and Microbial Habitability Categories and the Domain over Which Each Metric is Valid

| Metric Identifier | Microbial | Complex | Domain |
|---|---|---|---|
| $H^{W25}$ | Equations (1) and (2) | | Global |
| $H^{A25}$ | $273.15 \leqslant T_s \leqslant 373.15$ K and $P > 300$ mm $yr_{\oplus}^{-1}$ | ⋯ | Global |
| $H^{S08}$ | $273.15 \leqslant T_s \leqslant 373.15$ K | ⋯ | Global |
| $H^{S16}$ | ⋯ | $273.15 \leqslant T_s \leqslant 323.15$ K | Global |
| $H^{IF}$ | SIC $< 0.15$ | ⋯ | Marine |
| $H^{DG19H}$ | AI $> 0.39$ | ⋯ | Terrestrial |
| $H^{DG19NA}$ | AI $\geqslant 0.17$ | ⋯ | Terrestrial |

**Note.** $T_s$ represents surface air temperature, SIC sea ice concentration, and AI aridity index.
**References.** $H^{A25}$ (F. He et al. 2022; A. D. Adams et al. 2025); $H^{S08}$ (D. S. Spiegel et al. 2008; T. Jansen et al. 2019); $H^{S16}$ (L. Silva et al. 2016); $H^{IF}$(D. E. Sergeev et al. 2022); $H^{DG19H}$, $H^{DG19NA}$ (A. D. Del Genio et al. 2019).

### *3.2.2. Observed Habitability Calculation*

The global *observed* habitability $H^{obs}$ was calculated as the sum of the observed terrestrial and marine habitability, $H_T^{obs}$ and $H_M^{obs}$ respectively, which were in turn defined through the application of threshold values to annual mean NDVI ($NDVI_{mean}$) and Chl-*a* (Chl-$a_{mean}$), and additionally the annual minimum Chl-*a* (Chl-$a_{min}$):

$$H_T^{obs} = \begin{cases} \text{complex} & \text{if} \quad NDVI_{mean} > 0.3, \\ \text{microbial} & \text{if} \quad NDVI_{mean} > 0.15, \\ \text{limited} & \text{otherwise}, \end{cases} \quad (4)$$

$$H_M^{obs} = \begin{cases} \text{complex} & \text{if} \quad \text{Chl-a}_{min} > 0.15 \text{ mg m}^{-3}, \\ \text{microbial} & \text{if} \quad \text{Chl-a}_{mean} > 0.15 \text{ mg m}^{-3}, \\ \text{limited} & \text{otherwise}. \end{cases} \quad (5)$$

The microbial habitable threshold of the terrestrial (NDVI) data is taken at the maximum limit of the NDVI of bare ground, soil, sand, and rock, which typically show values of 0–0.15 (L. Phillips et al. 2008; R. Chatterjee et al. 2017; L. Wang et al. 2022). The threshold of 0.3 for terrestrial complex habitability is based on the average annual mean NDVI of all biomes excluding bare ground as described in R. S. Defries & J. R. G. Townshend (1994), as well as being the green threshold used in the sample MODIS NDVI figure (K. Didan 2021). For comparison, NDVI values near or below zero are indicative of bodies of water, snow, or ice (N. Pettorelli 2013; R. Chatterjee et al. 2017). Sparse vegetation, such as shrubs and grassland, typically show values of 0.2–0.5 while dense vegetation, including temperate and tropical forests, show values of 0.6–0.9 (R. S. Defries & J. R. G. Townshend 1994; N. Pettorelli 2013). Seasonal dynamics and phenology are also captured by NDVI, with woodland and forest areas falling to ≈0.2 during periods of senescence and high-latitude forests and tundra often showing low positive to negative values in the winter due to snowfall (R. S. Defries & J. R. G. Townshend 1994).

The marine complex and microbial thresholds are defined as where the annual minimum and mean Chl-*a* exceed 0.15 mg m$^{-3}$, respectively. This value was selected through the consideration of marine trophic states, which are commonly used to categorize aquatic environments in terms of nutrient availability and primary productivity, alongside the corresponding trophic levels that may be supported. Specifically, we examined the crossover from oligotrophic (nutrient poor with low productivity) to mesotrophic (moderate nutrients and productivity) environments, which coincides with a shift in abundance from almost entirely smaller primary producers (pico- and nanoplankton) to a mix of larger primary producers (nano- and microplankton) and primary consumers (micro- to mesozooplankton; T. Hirata et al. 2011; J.-N. Druon et al. 2019). The boundary between these two classifications in the open ocean is often defined in the range of 0.07–0.1 mg m$^{-3}$, although surface Chl-*a* in oligotrophic environments is known to seasonally exceed this range (D. Antoine et al. 1996; C. R. McClain et al. 2004; N. J. Hardman-Mountford et al. 2008; A. J. Irwin & M. J. Oliver 2009; A. Mignot et al. 2014; M. Cornec et al. 2021). Taking this into account, along with the slight positive bias seen at lower values in the OC-CCI Chl-*a* data set (Section 3.1), results in our selection of a slightly higher threshold value of 0.15 mg m$^{-3}$. The use of both annual minimum and mean Chl-*a* values acknowledges the short turnover time (≈2–6 days) of phytoplankton: the annual minimum exceeding the threshold implies the ability to continually support higher trophic states (i.e., complex life) year round, while the use of an annual mean recognizes the seasonal abundance of phytoplankton (i.e., microbial life; A. Longhurst 1995; C. B. Field et al. 1998). It is important to note here that the turnover time for terrestrial primary producers is around 19 yr, hence why it is sufficient to use only annual mean values for the NDVI threshold (C. B. Field et al. 1998). Although there exist other data sets representative of macroscopic life such as the biodiversity or species richness of marine mammals, these generally rely upon empirical observations, which are subject to data quality issues including spatiotemporal biases, gaps and errors, as well as differences in data collection methodology (M. D. Spalding et al. 2012; V. Moudrý & R. Devillers 2020). Furthermore, biodiversity metrics have been critiqued for inconsistency in their definition, with some ambiguity in their ecological value (D. S. Maier 2012). These points of contention can be avoided with the use of quantitative variables (such as NDVI or Chl-*a*), which are directly associated with biological productivity. In any case, the use of an annual minimum and mean Chl-*a* appears qualitatively well constrained with the declining abundance of higher trophic levels at the higher latitudes (S. Pompa et al. 2011; T. O. Gagné et al. 2020).

#### 3.2.3. Metric Validation

Each habitability metric was compared to the observed habitability by means of a qualitative spatial comparison, as well as quantitatively through computation of the chi-squared "test of independence" statistic ($\chi^2$), proportion correct or equivalently metric accuracy (PC), and the Heidke skill score (HSS; D. S. Wilks 2011). Computation of accuracy and skill scores is common to forecast verification in weather prediction (D. S. Wilks 2011; N. Boers et al. 2014; M.-L. Roussel et al. 2023; V. Torralba et al. 2023; J. Zhu et al. 2023), but has also been applied to GCMs (J. Kim & J. H. Ryu 2019; Y. Ryu et al. 2019; Y. Ma et al. 2023) and remote-sensing algorithms and data sets (A. Cheng et al. 2020; K. Yang et al. 2023). In particular, HSS measures the predictive power of the forecast relative to random chance, where a score of 0 implies that the forecast performs no better than random chance and a score of 1 implies a perfect forecast (D. S. Wilks 2011).

For each metric, relevant habitability category (microbial and/or complex), and domain (global, terrestrial, and/or marine), a contingency table was created of metric-generated "predicted" against observed habitability, from which these three statistics were computed. The contribution of each grid cell to the contingency table was weighted with respect to its area, such that the smaller grid cells toward the high latitudes receive a lower weight than those toward the equator. The combination of statistics allows for a more robust validation of each metric, whereby the proportion correct represents the metric accuracy, HSS attributes the accuracy to skill versus random chance, and $\chi^2$ is the statistical significance of the predicted and observed habitability belonging to the same population.

## 4. Results and Discussion

The broad spatial pattern of observed surface habitability shows complex habitability around the equator and midlatitudes, with microbial and eventually limited habitability assigned at higher latitudes and elevations (Figure 1(a)). Upon land, complex habitability also extends into the subtropical latitudes, with the exception of the Sahara desert and Himalaya. In comparison with the mean surface climate, habitability at lower latitudes is limited by areas of low precipitation ($P$) or negative net precipitation ($P - E$), higher latitudes limited by colder temperatures and resultant presence of sea ice, and microbial marine habitability coincides with the marginal ice zone (Figure A1). We also note that the lower-latitude belts of limited habitability approximately correspond to the extent of the Hadley cells (≈±30°N), indicative of the influence of planetary rotation rate upon the spatial patterns of surface habitability as suggested by other studies (A. D. Del Genio & R. J. Suozzo 1987; A. P. Showman et al. 2013; T. Jansen et al. 2019; A. D. Adams et al. 2025).

Comparing the observed habitability to the metrics based purely on surface air temperature ($T_s$), it is immediately apparent that the observed spatial pattern of habitability is not fully captured (Figures 1(a), (d), (e)). In particular, although the high latitudes are correctly labeled as limited habitability, lower-latitude areas such as the Sahara and marine oligotrophic gyres have been incorrectly labeled as habitable. Furthermore, the shift from microbial to limited habitability in $H^{S08}$ seems to more closely align to the latitudes at which the observed habitability switches instead from complex to microbial habitability, suggesting that the lower thermal limit of 0°C is a condition more representative of complex life. In the terrestrial domain, the validation statistics for these metrics indicate a statistically significant relationship with observed habitability and moderate metric performance with proportion correct (PC) ranging from 0.68 to 0.75 and HSS values of 0.33 to 0.36 (Table 2). Although still statistically significant, the temperature-based metrics show little predictive skill in the marine domain with an HSS of −0.03–0.06 and a low to moderate accuracy of 0.38–0.52, likely due to the incorrect categorization of the subtropical gyres. The lower performance in the marine domain also impacts the global results, where $H^{S08}$ shows a higher PC but lower HSS than $H^{S16}$. The $H^{IF}$ metric, which is defined using SIC, shows a similar issue of being unable to capture the low-latitude areas of limited habitability in addition to the limitation of only being applicable to the marine domain. The $H^{IF}$ validation results are also similar to the $H^{S08}$ marine values, due to the close relationship between the presence of sea ice and $T_s$ (J. A. Screen & I. Simmonds 2010).

The metrics defined by AI, $H^{DG19H}$ and $H^{DG19NA}$, are limited to the terrestrial domain through the definition of AI making use of PET (Equation (3)). Qualitatively, these metrics seem to better capture some of the areas of limited habitability seen at higher altitudes (including the Andes, Himalaya, and the Rockies) and more arid regions (including the Sahara and the Great Australian desert; Figures 1(g) and (h)). However, they respectively over- and underestimate the extent of limited habitability in these regions, as well as incorrectly assign the Greenland and Antarctic ice sheets as habitable. The validation statistics indicate better performance from $H^{DG19NA}$ with a PC of 0.70 and HSS 0.15, while $H^{DG19H}$ shows little predictive skill with HSS of −0.04 (Table 2). Despite a qualitatively better spatial representation of terrestrial habitability at lower latitudes, both AI-based metrics perform quantitatively worse than $H^{S08}$. To summarize, the metrics discussed so far are able to partially, but importantly not completely, describe the observed patterns of surface habitability seen on Earth. Furthermore, the patterns described by metrics dependent on $T_s$ are different to those dependent upon AI, and it may therefore be concluded that a metric definition should incorporate both elements in order to describe more wholly the patterns of surface habitability on Earth. In particular, the high HSS and PC values for the temperature-based metrics in the terrestrial domain indicate that $T_s$ is a good analog for determining climatological habitability, but is a less important factor in the marine domain. While adjusting the threshold for AI may improve the patterns of habitability captured in the terrestrial domain, it would be more beneficial to replace the use of AI entirely so as to remove the Earth-centric vegetation bias nested within the definition of PET as well as find a definition that would include the marine domain within predictions of habitable area.

Reviewing the validation results for the two metrics which contain conditions for both $T_s$ and water availability, namely $H^{A25}$ and our newly defined metric $H^{W25}$, we find an overall improvement relative to the other metrics across all domains for both microbial and complex habitability (Table 2). Qualitatively, $H^{W25}$ captures the global patterns of observed surface habitability most effectively, which is also reflected in the higher accuracies of 0.67 for microbial and 0.70 for complex habitability, which with higher HSS (0.34 and 0.36,

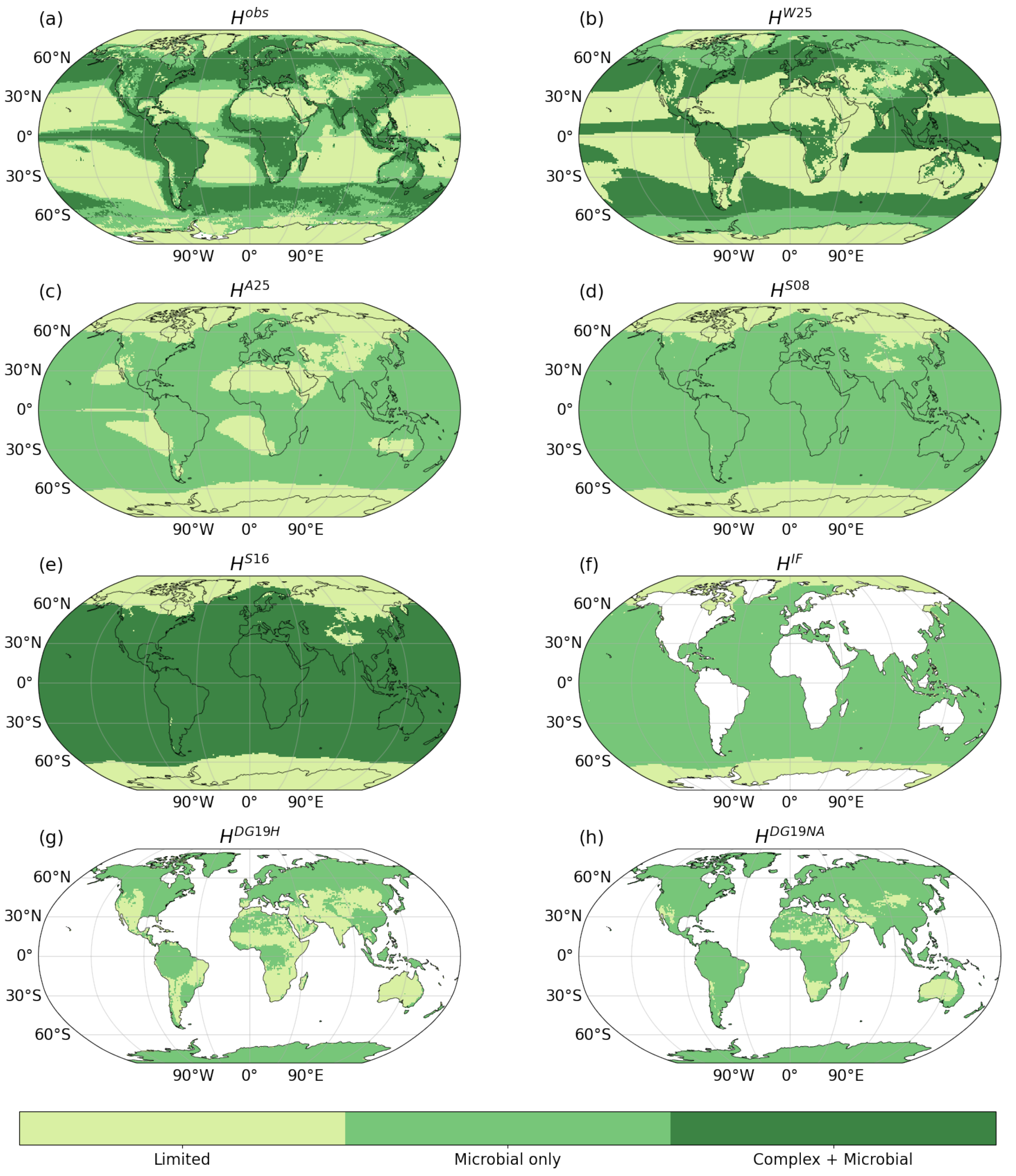

**Figure 1.** The observed and metric-derived climatological surface habitability of Earth based upon the annual mean climate across 2003–2018. The habitability criteria for each metric can be found in Table 1.

respectively) may be attributed to an improved predictive skill (Figure 1). Although only tested for the broader category of microbial habitability, $H^{A25}$ also shows a qualitative improvement globally, which seems to better capture the observed patterns of terrestrial habitability and to a lesser extent marine habitability. Again, this may be attributed to a modest increase in accuracy and predictive skill globally (PC = 0.61, HSS = 0.14) relative to its predecessor $H^{S08}$ (PC = 0.59, HSS = 0.05). However, similarly to $H^{S08}$ and $H^{IF}$, the 0°C lower limit in $H^{A25}$ also leads to a shift from microbial to

**Table 2**
Validation Statistics and Global Habitable Fraction for Each Metric and Domain against the Observed Habitability

| Domain and Metric | Microbial | | | | Complex | | | |
|---|---|---|---|---|---|---|---|---|
| | PC | HSS | $\chi^2$ | $f_H$ | PC | HSS | $\chi^2$ | $f_H$ |
| **Global** | | | | **0.59** | | | | **0.36** |
| $H^{W25}$ | **0.67** | **0.34** | **4906** | **0.53** | **0.70** | **0.36** | **5455** | **0.41** |
| $H^{A25}$ | 0.61 | 0.14 | 842 | 0.74 | … | … | … | … |
| $H^{S08}$ | 0.59 | 0.05 | 135 | 0.85 | … | … | … | … |
| $H^{S16}$ | … | … | … | … | 0.46 | 0.10 | 1333 | 0.85 |
| **Marine** | | | | **0.55** | | | | **0.29** |
| $H^{W25}$ | **0.64** | **0.28** | **2250** | **0.49** | **0.66** | **0.25** | **1898** | **0.40** |
| $H^{A25}$ | 0.52 | −0.02 | 21 | 0.81 | … | … | … | … |
| $H^{S08}$ | 0.52 | −0.03 | 83 | 0.89 | … | … | … | … |
| $H^{S16}$ | … | … | … | … | 0.38 | 0.06 | 671 | 0.89 |
| $H^{IF}$ | 0.53 | −0.02 | 31 | 0.90 | … | … | … | … |
| **Terrestrial** | | | | **0.71** | | | | **0.53** |
| $H^{W25}$ | 0.77 | 0.48 | 2881 | 0.62 | **0.80** | **0.60** | **4287** | **0.45** |
| $H^{A25}$ | **0.82** | **0.63** | **5229** | 0.56 | … | … | … | … |
| $H^{S08}$ | 0.75 | 0.36 | 1566 | **0.75** | … | … | … | … |
| $H^{S16}$ | … | … | … | … | 0.68 | 0.33 | 1615 | 0.75 |
| $H^{DG19H}$ | 0.52 | −0.04 | 22 | 0.60 | … | … | … | … |
| $H^{DG19NA}$ | 0.70 | 0.15 | 327 | 0.85 | … | … | … | … |

**Note.** Optimal metric values have been highlighted in bold. The observed fractional habitability for each domain is listed in bold in the row of each respective domain label. PC represents proportion correct (or equivalently, metric accuracy), HSS Heidke skill score, $f_H$ fractional habitability. Unless otherwise stated, $\chi^2$ scores have $p = 0.0000$ (to four decimal places) with 1 degree of freedom.

limited habitability at latitudes more representative of the complex-microbial boundary.

Both metrics perform particularly well in the terrestrial domain, with $H^{A25}$ showing the greatest improvement in performance and predictive skill for microbial habitability (PC = 0.82, HSS = 0.63) and $H^{W25}$ for complex habitability (PC = 0.80, HSS = 0.60). Large regions of limited habitability, including the Himalaya, Sahara, Greenland, and Antarctica are all correctly predicted by both metrics (Figure 1). There is a slight disagreement between $H^{W25}$ and the observed in the latitudes at which habitability shifts from complex to microbial, particularly on the Eurasian continent north of the Himalaya. In addition to the misaligned high-latitude shift to limited habitability, the microbial habitability seen along the northeastern coastline of Australia shown by $H^{A25}$ are more closely aligned to observed complex habitability. In comparison, Australia is not well captured by $H^{W25}$, which shows limited habitability across the region with some northern complex habitability, while the observed indicates microbial habitability across the Great Australian desert in addition to the northeastern complex habitability. The reduced habitability of the mountain ranges are somewhat captured by both metrics, although with some differences to both each other and the observed at a finer scale; for example, there is some disagreement in the spatial pattern of habitability across the Andes, and although the spatial patterns are similar in the Rockies, $H^{W25}$ underpredicts the habitability category as limited.

The differences between the metrics and the observed habitability in the terrestrial domain may be partly attributed to the assumptions made by the metrics. First, while an annual mean climate was used as an input—chosen for simplicity along with transferability to other planetary configurations such as tidally locked climates—its use ignores the seasonal nature of life on Earth. In particular, the assignment of microbial habitability in the northern latitudes on the Eurasian continent results from an annual mean $T_s$ between 253.15 K and 273.15 K. Although it holds that the northern temperate and boreal forests are unproductive during winter due to the suppression of photosynthesis at subzero temperatures, they are highly productive in the summer when no longer limited by temperature, resulting in the long-term accumulation of biomass (R. R. Nemani et al. 2003; M. Chen et al. 2017). Similarly, the seasonal snowfall in the higher latitudes can lead to negative values of NDVI during the respective hemispheric winters, in turn skewing the annual mean NDVI and the resultant observed habitability category (R. S. Defries & J. R. G. Townshend 1994). The high performance and predictive skill of both metrics indicate that $P$ is likely the primary controller of water availability in the terrestrial domain, and in particular, the improved accuracy and predictive skill of $H^{A25}$ relative to $H^{W25}$ suggests that $E$ may be an unnecessary input on land. Although the processes of evaporation and precipitation account for almost 90% of global water fluxes, the hydrologic cycle also includes many other terrestrial processes such as surface runoff and fluvial transport, glacial outflow, snow melt, and groundwater flow, as well as being subject to human influence (A. Dai & K. E. Trenberth 2002; K. E. Trenberth et al. 2007; R. P. Allan et al. 2020). While we focus here upon the atmospheric influence on surface water fluxes and highlight the simplicity in using solely $P$ (and additionally $E$ for $H^{W25}$), future work may wish to incorporate these other factors within future definitions of habitability, and investigate whether this improves the habitability categorization of the water-limited regions such as the mountain ranges and Australia. However, within the context of exoplanet habitability and current observational capabilities, this level of granularity may not be necessary given that both metrics are able to capture the broad-scale patterns of habitability (Y. Fujii et al. 2018).

In the marine domain, $H^{A25}$ exhibits little improvement to its predecessor $H^{S08}$ and despite beginning to show some limited habitability in the lower-latitude subtropical gyres, achieves comparable validation results (PC = 0.52, HSS = –0.02) which suggests that $P$ is not sufficient in describing water or nutrient availability (Figure 1, Table 2). In contrast, $H^{W25}$ shows a vast improvement in predictive skill and accuracy in comparison to the other metrics, for both microbial (PC = 0.64, HSS = 0.28) and complex (PC = 0.66, HSS = 0.25) habitability (Table 2). Furthermore, the latitudinal shift from complex to microbial habitability along with the limited habitability of subtropical gyres are qualitatively well captured by $H^{W25}$ (Figure A1). There is some disagreement in the assigned category of habitability in the high Arctic, which $H^{W25}$ predicts as microbial while the observed habitability almost directly shifts from complex to limited. However, in addition to the previously discussed low Sun-angle limitations of the OC-CCI Chl-*a* data set, satellite-derived data sets of Chl-*a* are also susceptible to pixel contamination introduced by nearby sea ice or mixed ice/ocean pixels and additionally show difficulty in detecting underice blooms (S. Bélanger et al. 2007; K. E. Lowry et al. 2014). In this way, it may be that OC-CCI Chl-*a* underestimates the true abundance of phytoplankton in the presence of sea ice, and accordingly the resultant observed habitability categorization. Another disagreement lies across the Indian and Western Pacific Oceans, whereby $H^{W25}$ predicts large-scale complex habitability while the Arabian Sea is assigned as limited. However, the observed habitability shows the inverse, with the Indo-Pacific Warm Pool typically showing low levels of productivity due to nutrient limitation, while the monsoonal winds in the Arabian Sea contribute toward a vertical mixing which, along with aeolian transport of iron fluxes, refreshes nutrient stocks and subsequently increases productivity (J. D. Wiggert et al. 2000; Y. Gao et al. 2001; T. K. Westberry et al. 2023). Similarly, the higher habitability observed, particularly along the western coast shorelines, which can be attributed to enhanced nutrient availability, has not been captured by our metric (L. A. Bristow et al. 2017; T. J. Browning & C. M. Moore 2023). Although showing considerable improvement relative to the other metrics, the $P - E$ condition is only able to partially capture the distribution of nutrients across the marine domain, and future work may aim to investigate using additional or alternate climate parameters to improve accuracy and predictive skill.

Finally, comparing the fractional habitabilities $f_H$ produced by the metrics demonstrates that the temperature-based metrics all overestimate $f_H$ in comparison to the observed values across all domains (Table 2). The AI-based metrics straddle the observed terrestrial microbial $f_H$, with the same pattern of under/overestimation respectively seen in $H^{DG19H}$ and $H^{DG19NA}$. Despite achieving the highest accuracy and skill score in the terrestrial domain, $H^{A25}$ predicts the lowest $f_H$ of all the metrics and underestimates $f_H$ by 0.15. In fact, the value is more comparable to the observed complex terrestrial habitability, suggesting that the $H^{A25}$ definition and in particular the $P$ condition is more descriptive of complex habitability, noting the similarity in definition to biome and biogeographical classifications (e.g., L. R. Holdridge 1947; P. T. Pennington et al. 2004). In the marine and global domains, $H^{A25}$ shows only minor improvement to its predecessor $H^{S08}$, with $f_H$ only marginally closer to the observed. Overall, $H^{W25}$ achieves $f_H$ most closely aligned to the observed values, with the exception of $H^{S08}$ in the terrestrial domain for microbial habitability, which is marginally closer to the observed $f_H$. However, the higher PC and HSS in $H^{W25}$ indicate an overall more accurate and skillful representation of terrestrial microbial habitability than $H^{S08}$ (Table 2).

To summarize, we have shown that the metrics that incorporate conditions for both temperature and water availability, $H^{W25}$ and $H^{A25}$, show a significant improvement of observed surface habitability across modern Earth in both qualitative and quantitative terms. In particular, $T_s$, $P$, and $E$ are all important factors in determining marine habitability, with $H^{W25}$ achieving accuracies of 0.64 and 0.66 in predicting microbial and complex habitability, respectively. $E$ seems to be a less important variable in the terrestrial domain, with $H^{A25}$ marginally outperforming $H^{W25}$ for microbial habitability. Simultaneously, $H^{A25}$ shows spatial patterns and a habitable fraction that may instead be more representative of complex habitability. Thus, for studies investigating surface terrestrial habitability, and in particular of macroscopic complex life (such as vegetation), it may therefore be recommended to use the definition provided by $H^{A25}$. For more general studies of surface habitability however, such as in the case of aquaplanet simulations (where continental configuration is unknown), or in parameter sweep studies (for example varying land fraction), we argue that using a single definition across both domains would offer a more consistent comparison, and in this case would recommend the use of $H^{W25}$.

This comparison makes use of mean surface climate variables, with the validation period taken here as an annual average across 2003–2018. Therefore, while these metrics aim to capture the mean climatological conditions that support surface life, we acknowledge that the climate on Earth exhibits variability over a variety of timescales, from the shorter-term diurnal and seasonal cycles, decadal to centennial modes of variability, to the longer-term Milankovitch cycles induced by orbital oscillations (M. Milankovitch 1941; A. S. von der Heydt et al. 2021). It is therefore important to also consider the rate of climate change when considering the evolution of surface habitability across time, particularly with respect to the time needed for the subsequent adaptation and evolution of organisms and ecosystems. For example, although the notably hothouse climates during the Permian–Triassic boundary and the Paleocene–Eocene Thermal Maximum could be classed as "more habitable" within temperature-based metric definitions, they coincided with mass extinction events as a result of sharp increases in surface temperature over relatively short time periods (H. Song et al. 2021). Similarly, there are other requirements of habitability that are not considered by these climate-based metrics (and are therefore assumed to be satisfied), including an active carbon cycle and plate tectonics, suitable atmospheric composition and pressure, as well as spectral energy distribution (J. F. Kasting et al. 1993; C. Cockell et al. 2016; R. M. Ramirez 2018, 2020). We therefore suggest climatologically defined metrics should be used in tandem with other measures of habitability, including but not limited to carbon cycle models, habitable zones and habitable zone lifetimes, modeling of "Gaia" or biosphere evolution, or nutrient availability (A. J. Rushby et al. 2013;

J. T. O'Malley-James et al. 2014; O. R. Lehmer et al. 2020; S. L. Olson et al. 2020; R. Arthur & A. Nicholson 2023; O. Herbort et al. 2024).

A second consideration regards the choice of validation data set, and in particular, our assumption of "photosynthetic life" acting as a proxy for "all surface life." While photosynthetic life is detectable from space and forms the basis of most surface ecosystems and therefore supports most heterotrophic life, the abundance and productivity of chemosynthetic life globally remains relatively underexplored and unconstrained (F. Baltar & G. J. Herndl 2019; A. E. Ray et al. 2022; F. Ricci & C. Greening 2024). However, the study of extremophiles has shown the ability for life on Earth to fill a wide range of environmental niches, and indeed, chemosynthetic life has been observed across a plethora of light-limited or anoxic environments that satisfy the three habitable requirements of liquid water, an energy source, and available nutrients, but may be considered unpalatable for photosynthetic organisms (C. Cockell et al. 2016; N. Merino et al. 2019). Most examples of these environments are subsurface, where chemosynthetic life supports whole ecosystems across hydrothermal vents, cold seeps, beneath ice shelves or glaciers, as well as within deep continental soils, sediments, and caves (E. S. Boyd et al. 2014; M. L. Pace et al. 2021; F. Ricci & C. Greening 2024). A growing area of research is exploring the presence of chemosynthetic life across extreme surface environments, including deserts, tundra soils, thermal springs, and Arctic freshwater lakes (D. Emerson et al. 2015; H. Šantrůčková et al. 2018; C. Greening & R. Grinter 2022; E. Magnuson et al. 2022; A. E. Ray et al. 2022; M. C. Fernandes-Martins et al. 2023; F. Ricci & C. Greening 2024). Within the upper layers of the ocean, chemosynthetic organisms and symbioses have been found to coexist among photosynthetic life, with abundances increasing with depth and inversely proportional to photosynthetic organisms (F. Baltar & G. J. Herndl 2019; E. M. Sogin et al. 2021; F. Ricci & C. Greening 2024). Given that all photosynthetic and many chemosynthetic organisms use the same biochemical pathway (i.e., the Calvin–Benson–Bassham cycle) for carbon fixation, they will share similar theoretical thermal limits, with a minimum of $\approx -25$°C needed for an active metabolism and a maximum of $\approx$140–150°C in which cellular structures begin to break down (C. Cockell et al. 2016; K. Harrison et al. 2025). We therefore maintain our assumption that if for some reason photosynthesis is unable to occur, for example on the nightside of a tidally locked planet or perhaps because the host star provides insufficient flux in the shorter wavelengths for efficient photosynthesis, we should expect that if life is to emerge, it could evolve to take advantage of the environment using an alternate metabolism such as chemosynthesis. With this in mind, along with our aim to minimize the introduction of additional biases by using an observational data set that measures photosynthetic reflectances for our validation, we do acknowledge there are likely biases in using a data set solely representative of photosynthesis. Future work may wish to explore other validation data sets, which may include empirical or modeled data but be more representative of other forms of life, such as species richness, biomass, or biological productivity (J. T. O'Malley-James et al. 2014; C. S. Cockell et al. 2021).

Finally, with many (exo)planetary modeling studies already making use of different climate-based definitions of surface habitability, we have aimed to constrain this definition to better reflect the observed patterns of habitability on Earth. In particular, the incorporation of $P$ and $E$ are advantageous over, for example, soil moisture or potential evapotranspiration in that they are predominantly driven by the atmospheric circulation and do not contain assumptions about vegetation (A. D. Del Genio et al. 2019; T. Jansen et al. 2019; A. D. Adams et al. 2025). While these variables (along with other influential parameters such as topography) are unlikely to be directly detectable on a subplanetary scale for the foreseeable future, the use of aquaplanet simulations may, to a first order, estimate surface habitability. This requires input variables that are either observable now (e.g., stellar flux and spectrum, planetary mass and radius, orbital period and eccentricity), or in the near future (e.g., atmospheric composition and for non-tidally locked planets, rotation period) with the Habitable Worlds Observatory and Large Interferometer For Exoplanets missions (S. Seager et al. 2005; A. D. Del Genio et al. 2019; S. P. Quanz et al. 2022). In the meantime, the existence and potential composition of an exoplanetary atmosphere may be estimated using currently available observables. First, the ability for a planet to retain an atmosphere may be assessed using the "cosmic shoreline" concept, which relates escape velocity (dependent upon planetary mass) with incident stellar flux (K. J. Zahnle & D. C. Catling 2017). Then, taking the assumption that the planet and its host star formed within the same protoplanetary disk, the bulk rocky composition of a planet may be inferred from the chemical composition of its host star (D. M. Jorge et al. 2022; R. J. Spaargaren et al. 2023). In tandem with planetary mass and radius, then the interior structure, outgassing efficiency and composition, and interior–atmosphere exchange may be estimated, providing constraints upon the resultant atmosphere (F. Gaillard & B. Scaillet 2014; L. Noack et al. 2014, 2017; R. J. Spaargaren et al. 2020; C. Brachmann et al. 2025). Although a connection between the abundance of volatiles (such as water) and orbital distance has been proposed, initial volatile budgets are typically varied due to the dependence on other processes relating to formation, migration, and evolution, enabling an exploration into the diversity of outgassed atmospheres (e.g., F. Albarède 2009; T. Yoshizaki & W. F. McDonough 2021; P. A. Sossi et al. 2022; C. Brachmann et al. 2025). Thus, a range of possible atmospheric compositions may be inferred from the stellar abundances and bulk planetary parameters, which may then inform GCM simulations exploring the climates and potential habitability in preparation for future atmospheric characterization.

## 5. Conclusions

We have reviewed a set of climate-based metrics of surface habitability against the spatial distribution of photosynthetic life on modern Earth, finding that a metric defined using either surface air temperature $T_s$, AI, or SIC alone is not sufficient in capturing the observed patterns of habitability. Metrics dependent upon $T_s$ and SIC approximately capture the shift to microbial and limited habitability at higher latitudes, but incorrectly label lower-latitude regions of observed limited habitability regions as habitable, resulting in an overestimation in the fractional habitability $f_H$. The metrics defined using AI show different patterns of habitability in comparison to those defined using $T_s$ and begin to resemble some of the lower-latitude terrestrial regions of limited habitability but exhibit poor predictive skill overall. We find that the metrics $H^{W25}$ and $H^{A25}$, which incorporate conditions for both temperature and water availability, provide the most complete representation of observed surface habitability across all domains. Although underestimating the respective fractional habitability, $H^{A25}$ shows the highest accuracy and skill score in the terrestrial domain for microbial habitability, indicating that the use of precipitation $P$ is an important factor in determining terrestrial habitability. Across all other domains and habitability categories, our metric $H^{W25}$, which is defined using $T_s$, $P$, and evaporation $E$, most closely aligns with the observed patterns both qualitatively and quantitatively. In particular, the vast improvement seen in the marine domain suggests that both $P$ and $E$ are valuable within habitability definitions. While the validation results of $H^{A25}$ imply that $E$ is not an essential factor in determining terrestrial habitability, we suggest that using a single definition across the global domain is advantageous for consistency, for example within aquaplanet simulations, or in other words, when continental configuration is unknown and so where the spatial patterns of surface habitability may be predicted to a first order by the atmospheric circulation. Globally, our metric shows accuracies of 0.67 and 0.70 for microbial and complex habitability, respectively, and it performs particularly well in the terrestrial domain with respective accuracies of 0.77 and 0.80. Disagreements between our metric and the observed can be attributed to a mixture of uncertainties from the data sources used to calculate observed habitability, seasonal variability, as well as the metric failing to entirely capture the distribution of nutrients across the marine domain. Future work may wish to explore the incorporation of additional climate parameters in order to better capture the global distribution of nutrients and surface water fluxes, as well as using other data sets for the validation that directly represent nonphotosynthetic life. However, we note that the metric in its current form offers a good representation of the observed surface habitability of modern Earth, with a dependence upon only three parameters commonly output by GCMs that are closely tied to the global atmospheric circulation and enable a first-order approximation of surface habitability upon exoplanets.

## Acknowledgments

The authors give thanks to Chris Brierley, Ian Crawford, and Rob Spaargaren for their insightful discussions and constructive feedback, as well as to the anonymous reviewers for their thoughtful suggestions which helped to shape this work. H.L.W. acknowledges support from NERC through the London NERC Doctoral Training Partnership (grant NE/S007229/1). N.J.M. acknowledges support from a UKRI Future Leaders Fellowship (grant MR/T040866/1) and the Leverhulme Trust through a research project grant (RPG-2020-82).

*Software:* cartopy (P. Elson et al. 2020), cftime (J. Whitaker et al. 2022), jupyter (T. Kluyver et al. 2016), matplotlib (J. D. Hunter 2007), numpy (C. R. Harris et al. 2020), regionmask (M. Hauser et al. 2024), scipy (R. Gommers et al. 2023), xarray (S. Hoyer & J. Hamman 2017), xesmf (J. Zhuang et al. 2023)

## Author Contributions

H.L.W. came up with the initial research concept and was responsible for executing the project and writing, editing, and submitting the manuscript. A.J.R and N.J.M. provided supervision for the project and feedback on multiple drafts of the manuscript.

## Appendix A
## Additional Figure

Figure A1 shows the mean surface climate of Earth across 2003–2018, with the variables shown corresponding to the habitability metric definitions outlined in Table 1.

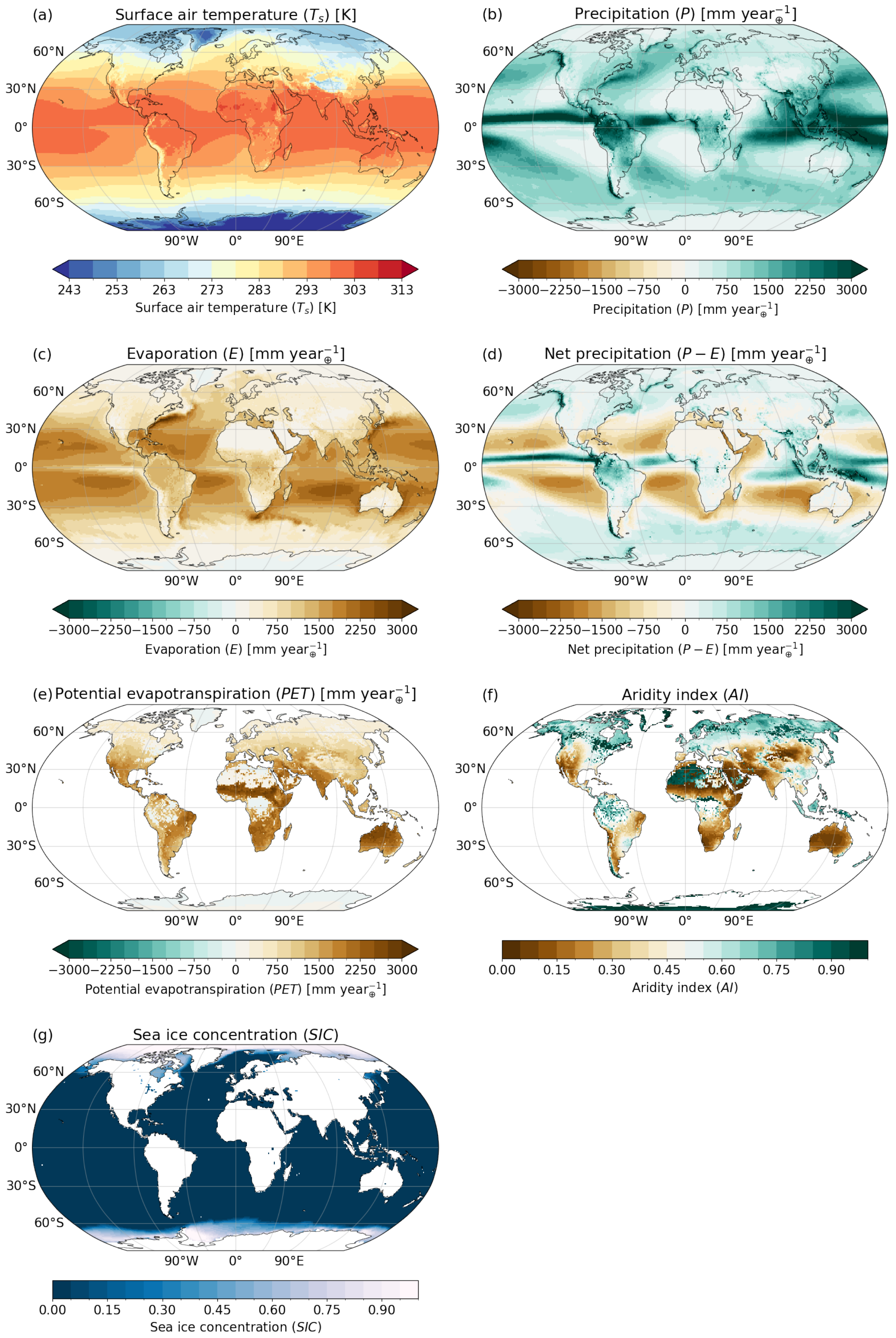

**Figure A1.** Annual mean climate across 2003–2018 derived from ERA5 for (a) surface air temperature $T_s$ (kelvin), (b) precipitation $P$ (mm yr$_\oplus^{-1}$), (c) evaporation $E$ (mm yr$_\oplus^{-1}$), (d) net precipitation $P - E$ (mm yr$_\oplus^{-1}$), (e) potential evapotranspiration PET (mm yr$_\oplus^{-1}$), (f) aridity index AI, and (g) sea ice concentration SIC.

## Appendix B
## Data and Code Availability

The project codebase is available on Zenodo (doi:10.5281/zenodo.15870061) and on GitHub.[6] The ERA5 monthly data set (H. Hersbach et al. 2023) is available for download from Copernicus Climate Change Service (2023). MODIS Terra NDVI (MOD13C2 v061; K. Didan 2021) is available for download at doi:10.5067/MODIS/MOD13C2.061. The OC-CCI monthly Chl-a ocean color data set (v6.0; S. Sathyendranath et al. 2019) is available for download at https://www.oceancolor.org. HadISST monthly sea ice concentration (N. A. Rayner et al. 2003; H. A. Titchner & N. A. Rayner 2014) is available for download at https://www.metoffice.gov.uk/hadobs/hadisst.

## ORCID iDs

Hannah L. Woodward https://orcid.org/0000-0002-9552-7918
Andrew J. Rushby https://orcid.org/0000-0001-6233-4311
Nathan J. Mayne https://orcid.org/0000-0001-6707-4563

[6] https://github.com/hannahwoodward/2025-hab-metrics